# Biomechanical modeling and computer simulation of the brain during neurosurgery

K. Miller[1,*], G. R. Joldes[1], G. Bourantas[1], S.K. Warfield[2], D. E. Hyde[2], R. Kikinis[3,4,5] and A. Wittek[1]

[1]Intelligent Systems for Medicine Laboratory, Department of Mechanical Engineering, The University of Western Australia, 35 Stirling Highway, Perth, WA 6009, Australia
[2]Computational Radiology Laboratory, Department of Radiology, Boston Children's Hospital and Harvard Medical School, 300 Longwood Avenue, Boston MA 02115
[3]Surgical Planning Laboratory, Brigham and Women's Hospital and Harvard Medical School, 45 Francis St, Boston, MA 02115
[4]Medical Image Computing, University of Bremen, Germany
[5]Fraunhofer MEVIS, Bremen, Germany

## Abstract

Computational biomechanics of the brain for neurosurgery is an emerging area of research recently gaining in importance and practical applications. This review paper presents the contributions of the Intelligent Systems for Medicine Laboratory and it's collaborators to this field, discussing the modeling approaches adopted and the methods developed for obtaining the numerical solutions. We adopt a physics-based modeling approach, and describe the brain deformation in mechanical terms (such as displacements, strains and stresses), which can be computed using a biomechanical model, by solving a continuum mechanics problem. We present our modeling approaches related to geometry creation, boundary conditions, loading and material properties. From the point of view of solution methods, we advocate the use of fully nonlinear modeling approaches, capable of capturing very large deformations and nonlinear material behavior. We discuss finite element and meshless domain discretization, the use of the Total Lagrangian formulation of continuum mechanics, and explicit time integration for solving both time-accurate and steady state problems. We present the methods developed for handling contacts and for warping 3D medical images using the results of our simulations. We present two examples to showcase these methods: brain shift estimation for image registration and brain deformation computation for neuronavigation in epilepsy treatment.

**Key Terms**: brain biomechanics, neurosurgical simulation, neuroimage registration, brain shift, meshless methods, image warping, epilepsy surgery, glioma surgery.

*Corresponding author:

Phone: +61 8 6488 8545; Fax: +61 8 6488 1024; E-mail: karol.miller@uwa.edu.au.

# 1 Introduction

By augmenting the surgeon's ability to perform operations, computer integrated surgery systems can increase surgical accuracy, improve the clinical outcomes and the efficiency of healthcare delivery. In this article we discuss the application of computational mechanics in computer-integrated neurosurgery systems. We discuss the physical and mathematical modeling approaches as well as the numerical solution methods used to solve the models developed at Intelligent Systems for Medicine Laboratory, The University of Western Australia and tested in collaboration with Harvard Medical School affiliated hospitals. We explain how to apply these methods in the areas of neurosurgical simulation and neuroimage registration.

## 1.1 Neuroimage registration

Many new therapeutic technologies, including stimulators, focused radiation, lesion generation, nano-technological devices, robotic surgery and robotic prosthetics have extremely localized areas of therapeutic effect [1]. Preoperative medical images capture only the preoperative (undeformed) anatomy of the patient. The ability to compute organ deformation and predict the intraoperative anatomy location is very important for performing reliable surgery on soft organs. In the context of image-guided neurosurgery, the surgeon must be informed about the location of pathologies and critical healthy areas in the brain during surgery. Brain displacements computed during surgery can be used to perform image registration, by warping preoperative high-quality images to the current, intra-operative configuration of the brain.

From a biomechanical point of view, the neuro-image registration problem involves large deformations, nonlinear material properties and nonlinear boundary conditions, and also requires patient-specific computational models. However, it differs from the previously discussed surgical simulation problem in two important ways: it requires accurate computations of the displacement field only (accurate reaction force and consequently precise stress computation is not important) and the computations must be conducted intra-operatively, with the results available to an operating surgeon in less than 40 s [2-5]. This still forms a stringent requirement for the computational efficiency of methods used.

## 1.2 Neurosurgical simulation

Neurosurgical simulators – software and visualization systems that simulate various aspects of brain surgery – have multiple medical applications, such as operation planning, simulation-based training and skill assessment. These applications require the modeling and simulation of the brain as a soft material interacting with surgical tools and the surrounding skull.

While simulations for operation planning can be performed offline, surgical simulation systems for surgeon training may have very stringent timing requirements in terms of visual and haptic feedback. These systems must compute the deformation field within a soft organ and the interaction force between surgical tools and tissue at frequencies of tens of Hz for visual feedback and at least 500 Hz for haptic feedback [6]. In order to satisfy these requirements, as well as to provide realistic simulation results, neurosurgical simulation systems require very efficient solution algorithms, capable of handling large deformations, nonlinear material properties and nonlinear boundary conditions.

Surgical training simulators can use a generic brain model, constructed based on population average organ geometry and material properties. However, simulations intended for operation planning must use patient specific computational models, which adds to the difficulty of the problem.

In Table 1 below summarizes requirements and our recommendations for modeling and computer simulation method choices, depending on the application. These recommendations are discussed in detail in this paper.

The paper is organized as follows: in Section 2 we describe our approaches to modeling geometry, boundary conditions, loading and material properties of the brain; in Section 3 we discuss numerical algorithms devised to efficiently compute the solution for the biomechanics-based brain models; in Section 4 we consider an important problem of how to use computed deformation fields to warp images; in Section 5 we present example applications; and finally in Section 6 we conclude with some reflections on the current state and perspectives for intraoperative use of patient-specific neurosurgical simulation.

**Table 1.** Summary of requirements and recommendations for brain biomechanics modeling depending on application, based on the Intelligent Systems for Medicine Laboratory's (ISML) 23 years of experience.

| | Type of Application | | |
|---|---|---|---|
| | **Intraoperative neuroimage registration** | **Neurosurgical simulation for training and/or skill assessment** | **Neurosurgical simulation for operationplanning** |
| **Patient specific?** | Yes | No | Yes |
| **Model components** | Brain parenchyma (separate treatment of gray and white matter not necessary), CSF (especially in ventricles), tumor. Explicit modelinf of the brain vasculature — not needed | Depends on specific application. | Depends on specific application. |
| **Real time?** | Yes, ca. between 40 s and 90 s to compute deformations and warp image. Practical limit can be defined as the time a neuro-surgeon is willing/can wait for the modeling results. | Yes, haptic rate 500 Hz or better | Yes, haptic rate 500 Hz or better |
| **Source of geometric information** | Anatomical MRI | Electronic brain atlas | Anatomical MRI |
| **Rapid generation of computational grid** | Needed | Not needed | Needed |
| **Spatial discretization method** | Meshless | Finite elements | Meshless |
| **Reliable computation of displacements** | Needed | Needed | Needed |
| **Reliable computation of internal stresses** | Not needed | Not needed | Not needed |
| **Reliable computation of reaction forces (e.g. acting on surgical tools)** | Not needed | Needed | Needed |
| **Loading** | Enforced motion of the exposed brain surface (measured intraoperatively) | Enforced motion of the boundary (at the contact with the virtual tool) | Enforced motion of the boundary (at the contact with the virtual tool) |

**Table 1.** (continuation from the previous page). Summary of requirements and recommendations for brain biomechanics modeling depending on application, based on the Intelligent Systems for Medicine Laboratory's (ISML) 23 years of experience.

| | Type of Application | | |
|---|---|---|---|
| | **Intraoperative neuroimage registration** | **Neurosurgical simulation for training and/or skill assessment** | **Neurosurgical simulation for operation planning** |
| **Constitutive model of intracranial constituents** | The simplest model compatible with finite deformation solution procedure: Neo-Hookean | Deformation gradients computed using the simplest model (Neo-Hookean) compatible with the finite deformation procedure are resubstituted to realistic constintutive model (we recommend Ogden-type) in the vicinity of the surgical tool where the reactions acting on surgical tool need to be computed. | Deformation gradients computed using the simplest model (Neo-Hookean) compatible with the finite deformation procedure are resubstituted to realistic constintutive model (we recommend Ogden-type) in the vicinity of the surgical tool where the reactions acting on surgical tool need to be computed. |
| **Brain-skull interaction model** | Frictionless finite sliding with separation | Frictionless finite sliding with separation | Frictionless finite sliding with separation |
| **Solution procedure** | Fully geometrically and materially nonlinear | Fully geometrically and materially nonlinear | Fully geometrically and materially nonlinear |
| **Time stepping** | Explicit with dynamic relaxation, the simplest central differences | Explicit, time-accurate, the simplest central differences | Explicit, time-accurate, the simplest central differences |
| **Practical model size** | 40-100 x $10^3$ degrees of freedom | 40-100 x $10^3$ degrees of freedom, more can be used for visualization. | 40-100 x $10^3$ degrees of freedom for computational grid, more can be used for visualization. |
| **Solution algorithm verification** | Against a converged finite element solution using hybrid (i.e. displacement-pressure formulation) finite elements. | Against a converged finite element solution using hybrid (i.e. displacement-pressure) finite elements. | Against a converged finite element solution using hybrid (i.e. displacement-pressure formulation) finite elements. |
| **Model validation** | Against intraoperative imaging, preferable intraoperative MRI | N/A. However, expert user feedback may be useful. | N/A. However, expert user feedback may be useful. |

## 2  Modeling approaches

The computation of brain deformation using the biomechanics-based approach requires the construction of a physical model as well as a mathematical model that can capture with sufficient accuracy the behavior of interest. The physical model includes the geometry, loading, boundary conditions and materials. Our approach to handling these different modeling aspects is described in this Section.

### 2.1  Geometry creation

The construction of a biomechanical model for computing brain deformation requires detailed geometric information, more specifically the surfaces representing the boundaries of the brain, the ventricles and other anatomical structures of interest (for example tumors). Such information can be extracted from brain atlases [7, 8] for applications that do not require patient-specific geometry (e.g. neurosurgical simulators for education, training and skill assessment). An example of a web-based electronic brain atlas is given in Figure 1.

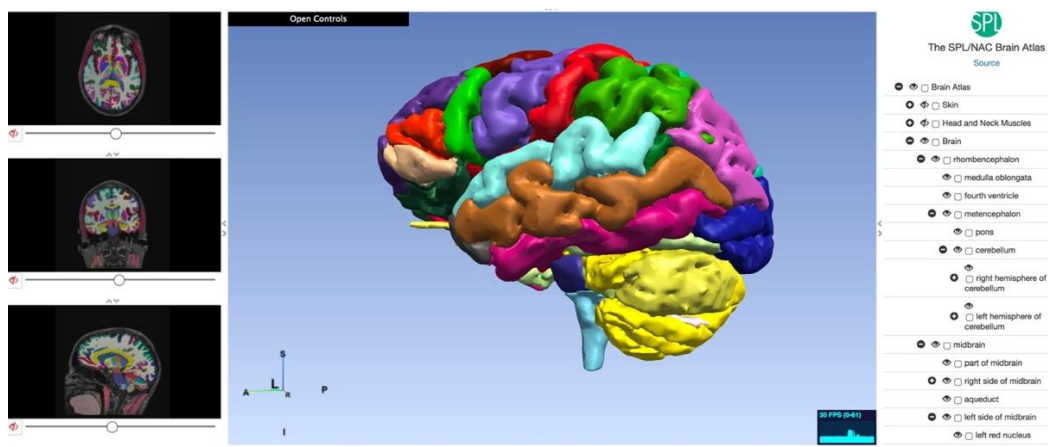

**Figure 1**. Multi-modality MRI-based Atlas of the Brain [9].

However, applications such as neurosurgical simulators for operation planning and image registration require patient-specific data, which can only be extracted from radiological images (such as MRI).

Voxel size in high-quality diagnostic preoperative MR images is usually of approximately $1mm^3$ magnitude. Therefore, patient-specific models of the brain geometry can be built with ~ 1 mm accuracy. It is usually considered that the precision of manual neurosurgery is at best 1 mm. Therefore, better than ~ 1 mm accuracy of brain model geometry is probably not required.

Brain image segmentation is usually used as a preprocessing step for geometry reconstruction. Despite recent advances [10-12], for most analysts it is a difficult, at best semi-automatic procedure. While the external brain surface can be relatively easily extracted using atlas-based segmentation (skull stripping), the internal brain structures are much harder to segment automatically, especially in the presence of pathologies such as tumors. For such cases, we propose the use of solution methods that do not require a segmentation of the internal structures, including meshless discretization and fuzzy tissue classification [13]. In Figure 2 we show a result of geometry preparation for a typical analysis of brain deformations for neuronavigation in cerebral glioma surgery.

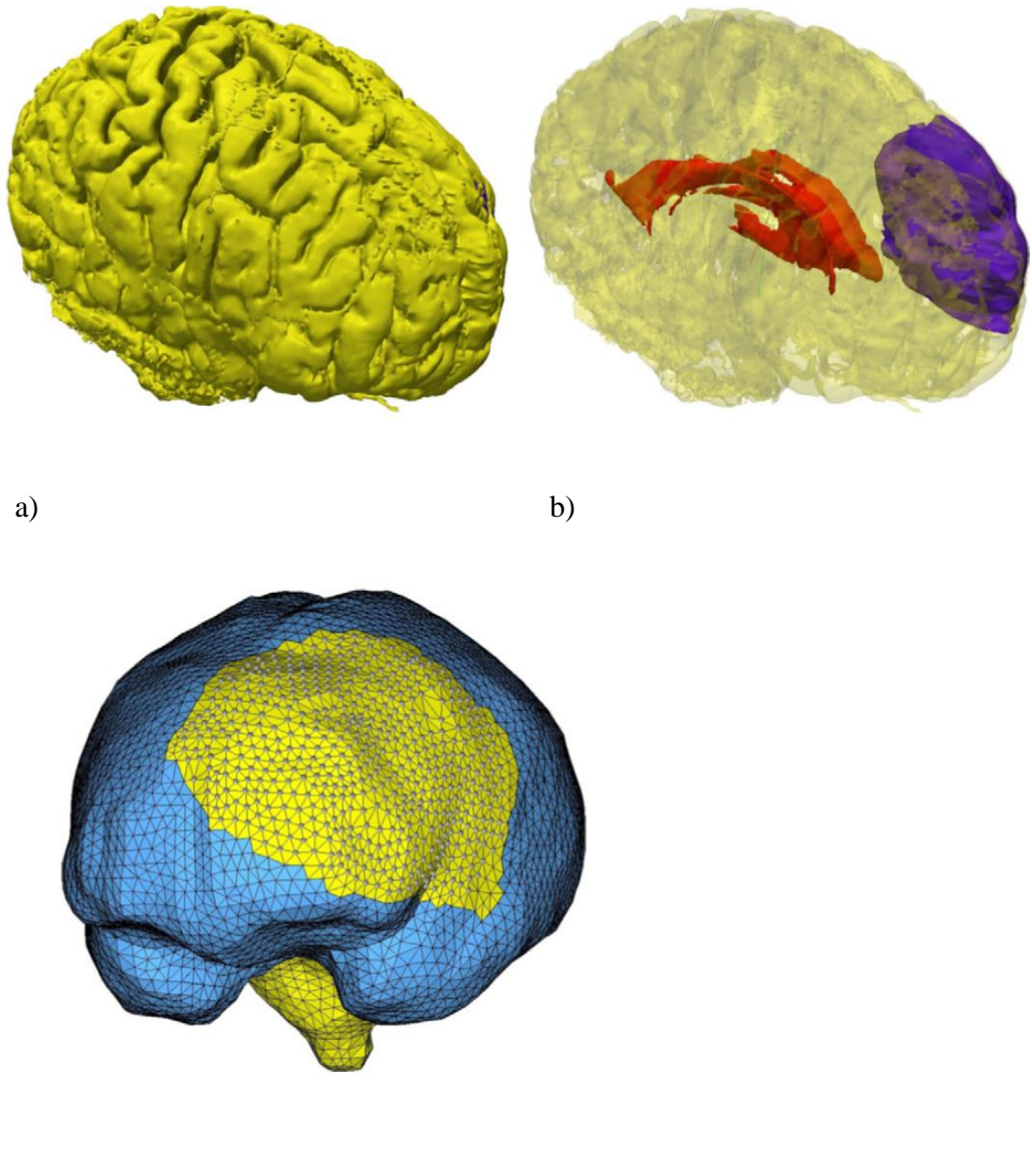

a)  b)

c)

**Figure 2**. Visualization of the problem geometry extracted from diagnostic (preoperative) MRI. a) Brain surface after skull stripping; b) Transparent view showing parenchyma (yellow), tumor (purple) and ventricle (red); c) Discretized (triangulated) skull surface (blue). The parenchyma surface (yellow) is visible through the craniotomy. Part of the brain stem (yellow) is also visible where it protrudes from the base of the skull. Image modified from Miller et al. [14].

*2.2 Loading and boundary conditions*

The accuracy of deformation computation from biomechanical models is influenced by the choice of appropriate and realistic loading and boundary conditions. The mathematical description of boundary conditions for neurosurgical simulation constitutes a significant problem because of the complexity of the brain–skull interface. The accurate measurement and quantification of the brain–skull interface behavior is currently an active research area [15-17].

Some researchers consider the brain surface to be fixed to the skull [18, 19]. Our experience [5, 20] and the study by Hu et al. [21] suggest the presence of relative movement between the brain and the skull. Therefore we model the brain-skull interface as a frictionless contact which allows separation. The appropriateness of this choice was further confirmed by an extensive parametric study of Wang et al. [22].

The skull is much stiffer than the brain tissue, and therefore it is assumed to be rigid. The constraining effects of the spinal cord on the brain rigid body motion and spine–spinal cord interactions and can be simulated by constraining the spinal end of the model.

We advocate the definition of loading for the biomechanical models as imposed displacements on the model's surface [5, 23, 24]. When modeling the deformation of the brain due to its interaction with a surgical tool, the loading will be defined by the known motion of the surgical tool. When performing intraoperative image registration, the motion of the exposed part of the brain surface can be measured using a variety of techniques (such as stereo vision, intraoperative MRI or ultrasound) [25] and then be used to define the model loading.

When loading is prescribed as imposed motion on the boundary, the computed deformation field within the domain depends very weakly on the mechanical properties of the continuum [23, 24, 26]. Given the great uncertainty regarding patient-specific properties of the brain tissue and the large variability between patients [27, 28], this feature is of great importance in the biomechanical modeling of the brain.

*2.3 Material properties*

As evidenced by our numerous experimental studies, the mechanical response of brain tissue to loading is very complex [29, 30], with a highly nonlinear stress–strain relationship and with tissue stiffness in compression much higher than in extension. There is also a nonlinear relationship between stress and strain rate, with stresses at a moderate strain rate (0.64 $s^{-1}$) being about ten times higher than at the low strain rate of 0.64x$10^{-5}$ $s^{-1}$. These results are in general agreement with measurements conducted by other groups [31-35]. Despite this complexity, for scenarios where modeling results are weakly dependent on the material properties (as described in the previous subsection), we use a Neo-Hookean material model [13, 36-38], the simplest model suitable for modeling large deformation behavior.

Neurosurgical simulations often involve computation of reaction forces acting on surgical tools. For such simulations, we recommend computing deformation field within the brain using the Neo-Hookean model and prescribing the loading through imposition of the known motion of the surgical tool. Realistic stresses in the vicinity of the tool are then computed by resubstituting the deformation gradients in the vicinity of the tool onto a more realistic constitutive model. We recommend an Ogden-based hyperviscoelastic constitutive model, capable of capturing the highly nonlinear behavior of the brain tissue [39]. Neo-Hookean and Ogden-type models discussed here have the advantage that their implementation is readily available in commercial finite element software [40, 41]. Our simulations (Section 5) suggest that assigning different properties to gray and white matter does not appreciably improve the accuracy of the computed displacements.

The commonly applied material models assume the brain tissue to be incompressible and isotropic. The assumption of incompressibility is common for soft hydrated tissues [27, 28, 42-47]. In experiments on brain tissue at moderate strain rates we found this assumption to hold [48]. Very soft tissues, such as the brain, do not normally bear mechanical loads and do not exhibit directional structure (provided that a large enough sample is considered). Therefore, their behavior can be assumed to be isotropic [29, 32, 42, 47, 49-51]. Prange and Margulies [34] reported anisotropic properties of brain tissue, measured using very small samples, approximately 1 mm wide. At such a small scale, the fibrous nature of most tissues will cause detectable difference in directional properties. We believe that at the length scales relevant to models of surgical procedures (~ 1 cm), the assumption that the brain tissue does not exhibit any directional variation of mechanical properties is justified [30].

Because of the very large variability inherent to biological materials [29-31, 34, 39], average material properties determined based on population studies (often on animal tissues) are clearly not sufficient for patient-specific computation of stresses and reaction forces. Unfortunately, despite progress in non-invasive tissue characterization using ultrasound and MR elastography [30], reliable methods of measuring patient-specific properties of the brain at large strains are not yet available.

## 2.4 Mathematical model

Once the physical biomechanical model is created, we need to consider the governing equations that describe its behavior. We consider the brain deformation is described by the equations of continuum mechanics (balance of mass, linear momentum, angular momentum and energy), written in a Lagrangian framework. These partial differential equations, together with boundary and initial conditions, define the mathematical model associated with the biomechanical model.

# 3    Solution methods

The partial differential equations describing the behavior of the brain are too complex for an analytical solution. The only option is to use numerical solution methods and compute a numerical solution. In this Section we describe the main aspects of the numerical solution methods employed.

## *3.1    Continuum mechanics formulation*

The great majority of commercial finite element programs use the Updated Lagrangian formulation, where all variables are referred to the current configuration of the system (i.e. at the end of the previous time step). The advantage of this approach is the simplicity of incremental strain description, which simplifies implementation of some complex material behavior (such as plasticity). The disadvantage is that all derivatives with respect to spatial coordinates must be recomputed in each time step, because the reference configuration is changing.

We use the Total Lagrangian formulation [52], where all variables are referred to the initial (undeformed) configuration of the system. The main advantage of this formulation is that all derivatives with respect to spatial coordinates are calculated relative to the original configuration and therefore can be precomputed, which is particularly important for time-critical applications such as surgical simulation and intraoperative image registration. Furthermore, the hyperelastic or hyperviscoelastic brain material models are much easier implemented in the Total Lagrangian formulation, as these material models can be easily described using the deformation gradient, which is calculated relative to the undeformed configuration of the system.

## *3.2    Time integration*

The integration of continuum mechanics equations in the time domain can be performed using either implicit or explicit methods [53, 54]. Some of the implicit time integration methods are unconditionally stable, with the maximum time step that can be used limited only by accuracy considerations. However, implicit time integration methods require a set of nonlinear algebraic equations to be solved at each time step. Furthermore, multiple iterations may be needed at each time step to control the error and prevent divergence. Therefore, the number of numerical operations per each time step can be three orders of magnitude larger than in the case of explicit integration [53].

In explicit time integration methods, such as the central difference method, the treatment of nonlinearities is straightforward and no iterations are required. After discretization, the equations of motion can be decoupled for each degree of freedom through the use of a lumped (diagonal) mass matrix [53]. This eliminates the need for assembling the stiffness matrix of the entire model, leading to significantly reduced computational cost for each time step and internal memory requirements, as compared to implicit integration. These characteristics make explicit integration suitable for real-time applications.

A main disadvantage of explicit time integration methods is their conditional stability: the time step that can be used is limited by a maximum allowable time step, which ensures solution stability. Nevertheless, the stiffness of the brain tissue is very low (about eight orders of magnitude lower than that of common engineering materials such as steel) [30]. Since the maximum time step allowed for stability is inversely proportional to the square root of Young's modulus divided by the mass density [40], it is possible to conduct brain deformation simulations with much longer time steps than in typical dynamic simulations in engineering. Therefore, in our solution methods we combine the Total Lagrange formulation with explicit time integration. We use the Total Lagrange Explicit Dynamics (TLED) algorithm in the context of both finite element [52, 55] and meshless discretizations [56]. The TLED algorithm offers the following benefits: allows the precomputing of many variables of interest (e.g. derivatives with respect to spatial coordinates, hourglass control parameters); no accumulation of errors (increase stability for quasi-static solutions); easy implementation of the material law for hyperelastic materials using the deformation gradient; straightforward treatment of nonlinearities; no iterations required in each time step; no large system of equations needs to be solved (requires only vector computation); low computational cost for each time step; well suited for parallel implementation [57, 58].

*3.3 Steady state solution*

Some brain deformation problems, such as neuro-image registration, require the steady state solution of the continuum mechanics problem. We compute the steady-state solution using Dynamic Relaxation (DR). DR is an explicit iterative method which relies on the introduction of an artificial mass dependent damping term in the equation of motion, to attenuate the oscillations in the transient response, resulting in convergence towards the steady state solution [59, 60].

Being an explicit solution method, DR is well suited for solving highly nonlinear problems (including both geometric and material nonlinearities), having all the characteristics of an explicit solution method described in the previous subsection. For such problems the DR parameters that ensure fast convergence change during the course of the simulation. We have developed an adaptive method for computing the DR parameters values that increase the convergence speed for nonlinear deformation problems [60]. The adaptive parameter estimation method involves only vectors, preserving the computational advantages of the explicit DR method.

*3.4 Finite element discretization*

An important step in obtaining a numerical solution for the mathematical model of the brain deformation problem is spatial discretization – the creation of a computational grid [61]. In many cases the computational grid takes the form of a finite element mesh. For applications with stringent computational time requirements, such as surgical simulation or intra-operative image registration, it is desired that the mesh is constructed using low-order elements that are computationally inexpensive. We suggest the use of the linear under-integrated (i.e. with a single Gauss integration point) hexahedron.

While tetrahedral meshes can be generated automatically, automatic hexahedral mesh generation for complex geometries is still an unsolved problem [61]. Template-based meshing algorithms can be used for discretizing different organs using hexahedrons [62-65], but these types of algorithms have grave difficulties with pathological cases – the ones of greatest interest – as the position and size of a pathology (such as a tumor) cannot be predicted up front and therefore it's inclusion in the preprepared atlas is very difficult. The practical impossibility to automatically generate hexahedral meshes is one of the main reasons why many authors use tetrahedral meshes for their models [3, 66-68]. In order to automate the simulation process, mixed meshes having both hexahedral and linear tetrahedral elements are the most convenient (Figure 3), as a trade-off between the effort to generate patient-specific computational grid and the speed and accuracy of computing a solution.

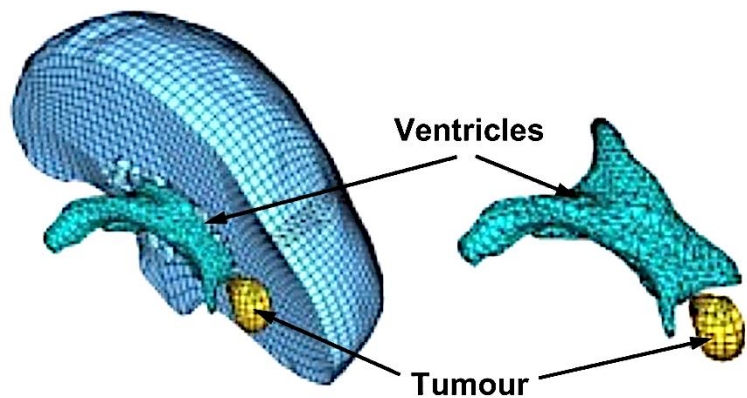

**Figure 3**. Typical example of a patient-specific brain mesh [69]. This mesh consists of 14,447 hexahedral elements, 13,563 tetrahedral elements and 18,806 nodes.

The under-integrated hexahedral elements are very efficient from a computational point of view, but require the use of an hourglass control algorithm in order to eliminate un-physical deformations due to zero energy modes, which are the artefact of grossly simplified, single-point integration [70]. Flanagan and Belytschko [70] proposed an effective hourglass control algorithm that has been implemented in a number of commercial finite element codes. Their method is applicable for hexahedral and quadrilateral elements with arbitrary geometry, undergoing large deformations described using updated Lagrangian formulation.

We extended this algorithm to the Total Lagrangian formulation [71]. In our approach all quantities except nodal displacements are constant and can be precomputed. This is a key feature of the Total Lagrangian formulation: most quantities used in calculations can be pre-computed in a preprocessing stage allowing a time-sensitive, intraoperative part of computation to be very efficient.

As fast, automatic generation of good-quality hexahedral meshes for objects with complicated geometry is not yet possible, our methods must cope with a substantial number of tetrahedral elements in the mesh, Figure 3. When modeling incompressible continua, linear tetrahedral elements exhibit artificial stiffening (often referred to as volumetric locking). This phenomenon occurs also for nearly incompressible materials and therefore introducing slight compressibility does not solve the problem.

Different authors proposed various improved linear tetrahedral elements with anti-locking features [72-75]. The average nodal pressure (ANP) tetrahedral element proposed in [72] is computationally inexpensive and provides much better results for nearly incompressible materials compared to the standard tetrahedral element. Nevertheless, one shortcoming of the ANP element and its implementation in a finite element code is the handling of interfaces between different materials. We extended the formulation of the ANP element so that all elements in a mesh are treated in the same way, requiring no special handling of the interface elements [76].

### 3.5 Meshless discretization

An alternative to the finite element method is to take advantage of one of the available meshless methods. The problem of computational grid generation disappears, as one needs only to drop a cloud of points inside the geometry being discretized [77-79], see Figure 4.

The motivation for the use of meshless methods is very compelling: they allow simple, automatic computational grid generation for patient-specific simulations. We use a modified Element-Free Galerkin (EFG) method that is meshless in the sense that shape functions are defined based on nodes without taking into account any nodal connectivity information. Node placement is almost arbitrary. Geometrically nonlinear Total Lagrangian formulation is used together with explicit time integration, which makes our meshless algorithm (MTLED [80, 81]) in many respects similar to the TLED algorithm used with the finite element method [56, 82].

In EFG the shape functions are created using the moving least-squares (MLS) approximation [83]. We have developed a modified version of the MLS approximation, which can use second order polynomial base functions under the same conditions (in regards to the nodal distribution) as for linear polynomial base functions, leading to higher approximation accuracy [84-87].

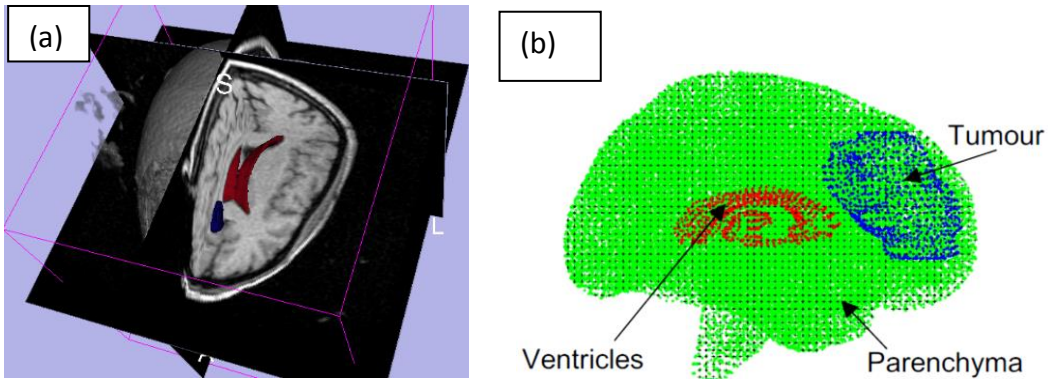

**Figure 4.** (a) 3D magnetic resonance image presented as a tri-planar cross-section, with over-imposed ventricles (red) and tumor (blue) segmentations. (b) 3-D patient-specific meshless discretization. The integration points are indicated as (+) and interpolating nodes — as (•) for the brain parenchyma, (•) for the ventricles, and (•) for the tumor. Regular hexahedral background integration grid was used [14].

The MLS shape functions are non-interpolating, which poses difficulties in the imposition of essential boundary conditions (EBC). In order to overcome these difficulties, we have recently developed a new method of enforcing the essential boundary conditions in EFG based meshless TLED framework, named Essential Boundary Conditions Imposition in Explicit Meshless (EBCIEM) [88]. The method computes displacement corrections which are added to the displacement field during time stepping to impose the displacements on the essential boundary. While these corrections involve some additional matrix computations in every time step, in the case of Total Lagrangian formulation these matrices are constant and only need to be computed once at the beginning of the simulation.

Gaussian quadrature over a background mesh is typically used for integration in EFG methods [56, 78, 89, 90]. There are two sources of error associated with these integration schemes [89]: the shape functions in EFG methods are not polynomial but rather rational functions (which are not really suitable for Gauss integration) and their local support may not align with the integration cells. Therefore, exact integration in EFG methods is practically impossible. It is also very difficult to evaluate how much error in the computed solution is due to inaccuracies in numerical integration. To solve these problems we have developed an adaptive numerical integration procedure for meshless methods [91, 92]. A function constructed based on the shape functions guides an adaptive integration cell division process which stops only when the guiding function is integrated with a prescribed accuracy. The resulting distribution of integration points and corresponding weights can then be used to compute all integrals involved in the solution of the problem. The use of Total Lagrangian formulation means that the shape functions do not change during the solution process, and therefore the distribution of integration points used for numerical quadrature may not need to be updated. Because integration accuracy can be controlled, the developed adaptive integration method can also be used for obtaining very accurate reference solutions and evaluate the influence of integration error for other numerical integration schemes.

*3.6 Brain-skull interaction*

Many biomechanics-based simulations require the treatment of interactions between different parts of the model, such as the contact between the brain and the skull. The brain-skull interaction can be treated as contact of a deformable object (the brain) and a rigid surface (the skull). We developed a very efficient algorithm that treats this interaction as a finite sliding, frictionless contact, which allows separation [93]. Unlike contacts in commercial finite element solvers (e.g. ABAQUS, LS-DYNA), our contact algorithm has no configuration parameters (as it only imposes kinematic restrictions on the movement of the brain surface nodes) and is very fast, with the speed almost independent of the mesh density for the skull surface.

The contact algorithm works by detecting the nodes on the brain surface (also called the slave surface) which have penetrated the skull surface (master surface) and displacing such slave nodes to the closest point on the master surface.

**4   Biomechanics-based image registration**

For biomechanics-based image registration, the computation of the deformation field using the biomechanical model is only the first step in performing the image registration. During surgery, the computed deformation field must be used to warp the preoperative image to the intraoperative organ configuration. As this must be done intraoperatively, the image warping algorithms must be very efficient. The image warping algorithm we developed is presented in the following subsection.

*4.1   3D image warping*

The deformation field predicted by the patient-specific biomechanical model using a finite element or meshless solution method can be seen as a forward transform $T$ which indicates the position in the target (deformed) image for all the points used for discretizing the source image (Figure 5) [94]. In order to create the target image, the origin of each of its voxels in the source image is required; therefore the backward transform $X_S = T^{-1}(X_T)$ needs to be computed. Once the position in the source image is found for a voxel, its intensity can be computed by interpolation.

| Target (deformed) image | | Source (undeformed) image |

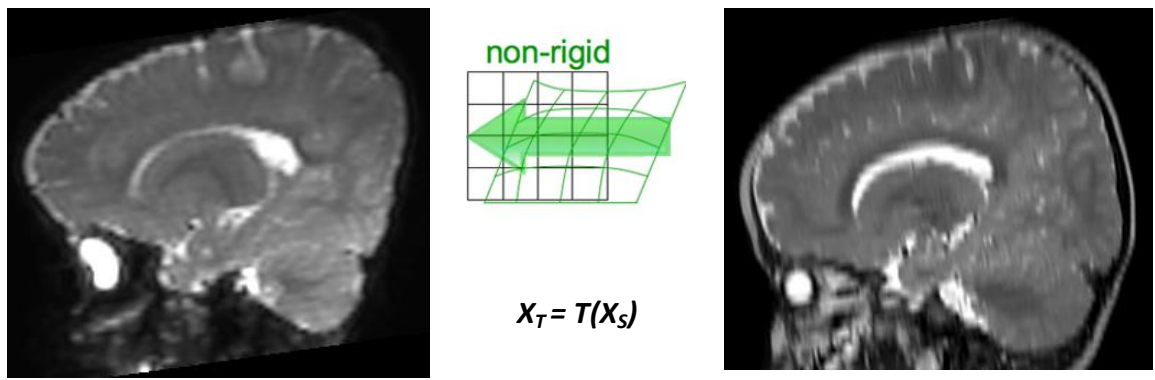

$X_T = T(X_S)$

**Figure 5.** The forward transform *T* defines the position in the target image $X_T$ for any point in the source image $X_S$; it is defined by the computed displacements at the nodes used for discretization.

Most image-based registration algorithms either use transforms which are easily invertible (rigid, affine) or the backward transform is computed directly by the registration procedure (B-Spline). When FEM or meshless solution methods are used, the transform is defined by the displacements computed at nodes. Therefore, evaluating the transform or its inverse at any other points (such as the centers of the voxels in the target image) will require spatial interpolation [95].

Because the positions of the nodes used for discretization in the target image space are not known until the solution of the biomechanical model is obtained, it is not possible to use a scattered data interpolation method that allows precomputation of some of its parameters; the entire spatial interpolation procedure needs to be performed for a very large number of voxels (even a relatively low resolution image can have 256*256*128 = 8,388,608 voxels).

Our initial approach to image warping [96] used the finite element mesh, in its deformed configuration, for performing the spatial interpolation. The method proved to be too slow for intraoperative image registration. Therefore, we developed a new method, which takes the deformation field computed by an FEM or meshless method and approximates it by a cubic B-Spline using a multi-level B-Spline scattered data interpolation approach [97]. We implemented the algorithm and released it as an extension to 3D Slicer, an open source software platform for medical image processing and three-dimensional visualization [98, 99]. The generation of the B-Spline approximation is very fast, and the computed transform can be used directly by existing image warping software, such as 3D Slicer, Figure 6.

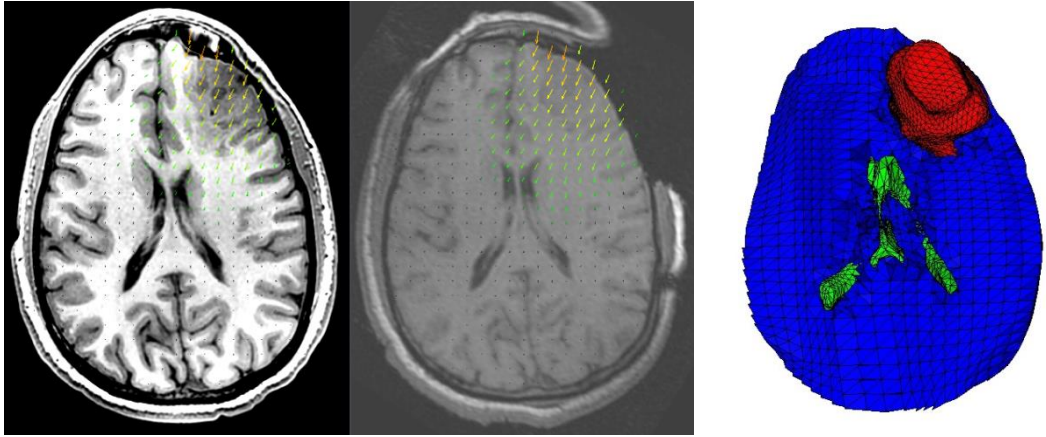

**Figure 6.** Craniotomy-induced brain shift for a case from Garlapati et al. [37]. The deformed high resolution preoperative image (left) is compared to the intraoperative image (center); the arrows show the size and direction of the brain tissue displacements. A section through the brain computational model used to predict the brain shift, showing the ventricles (green) and tumor (red) is presented on the right. The deformed image was obtained using the 3D Slicer extension ScatteredTransform [99].

## 5 Application examples

### 5.1 Craniotomy-induced brain shift simulation

A particularly exciting application of nonrigid image registration is in image-guided procedures, where preoperative scans are warped onto sparse intraoperative images during the operation, in real time [4, 100]. We are especially interested in registering high resolution preoperative MRIs with lower quality intraoperative imaging modalities, such as intraoperative ultrasound [101] and MRI [102]. To achieve accurate matching of these modalities, precise and fast algorithms to compute brain tissue deformations are fundamental.

In the following subsections we discuss outcomes of the analysis of 33 cases of craniotomy-induced brain shift representing different scenarios that may occur during an operation [69, 102].

#### 5.1.1 Generation of patient-specific computational grids from medical images

We selected preoperative and intraoperative medical image datasets of 33 patients with cerebral gliomas from a retrospective database of a very large number of intracranial tumor cases available at the Children's Hospital in Boston [103]. Imaging was performed using a 0.5T open MR system in the neurosurgical suite. The resolution of the images is $0.85 \times 0.85 \times 2.5$ mm$^3$. Consent was obtained for the use of the anonymized retrospective image database, in accordance with the Institutional Review Board of the Children's Hospital in Boston.

We created a three dimensional (3D) surface model of each patient's brain from segmented preoperative magnetic resonance images (MRI), see Figure 2a in Section 2.1 above. In our models, we assigned different material properties to the parenchyma, tumor and ventricles, and therefore we had to segment the parenchyma, ventricles and tumor. We used the region growing algorithm implemented in 3D slicer, followed by manual correction.

To meet the requirement of close-to-real time computations we chose low-order elements (8-noded hexahedron and 4-noded tetrahedron) which are inexpensive when used with our explicit solution methods [58]. To prevent volumetric locking, tetrahedral elements with average nodal pressure (ANP) formulation, were used [76]. The meshes were generated in two steps. First we applied meshing software from our collaborators at the University of Iowa IA-FEMesh [104]. Then we refined and corrected the meshes with HyperMesh (commercial FE mesh generator by Altair of Troy, MI, USA). Figure 3 (Section 3.4) shows a typical mesh, consisting of 14,447 hexahedral elements, 13,563 tetrahedral elements and 18,806 nodes.

### 5.1.2 Displacement loading and boundary conditions

We chose to load our models through forced displacements on the exposed (due to craniotomy) surface of the brain, Figure 2c. This is important as it allows treating the brain-shift computation as a Dirichlet-type problem [26]. This approach requires only replacing the brain-skull contact boundary condition with prescribed displacements, and therefore no mesh modification is required at this stage. Initially we aligned the preoperative and intraoperative coordinate systems by rigid registration. Then we determined the displacements at the mesh nodes located on the exposed surface of the brain with the interpolation algorithm from Joldes et al. [105].

As explained in Miller et al. [26] and Wittek et al. 106], for Dirichlet-type problems where loading is prescribed as forced motion of boundaries, the unknown deformation field within the domain depends very weakly on the mechanical properties of the continuum. This feature is of great advantage in biomechanical modeling for medicine where there are always uncertainties in patient-specific properties of tissues.

A frictionless sliding contact with separation has been chosen to model brain-skull interaction, see Section 3.6 above.

### 5.1.3 Mechanical properties of the intracranial constituents

The solution in displacements of Dirichlet-type problems is only weakly affected by the constitutive model of the brain tissue [26]. Therefore, we used Neo-Hookean model with Young's modulus of 3000 Pa [107] - the simplest hyperelastic constitutive law compatible with finite deformation (nonlinear) solution procedure. The Young's modulus of the tumor was assigned a value twice larger than that for the parenchyma, keeping it consistent with the experimental data of Sinkus et al. [108]. This admittedly fairly arbitrary choice has little influence on the computed displacement field as evidenced by our parametric study [109].

As the brain tissue is almost incompressible, we chose Poisson's ratio of 0.49 for the parenchyma and tumor [5]. Brain tissue's incompressibility necessitates the use of non-locking tetrahedral elements such as the one described in Joldes et al. [76, 110]. The ventricles were treated as a very soft compressible hyperelastic solid with Young's modulus of 10 Pa and Poisson's ratio of 0.1 [5].

### 5.1.4 Solution algorithm

For the time integration of partial differential equations of equilibrium we used the explicit method described in Section 3 above. When a diagonal (lumped) mass matrix is used, the discretized equations are decoupled. Therefore, even for nonlinear problems, no matrix inversions and iterations are required. Application of the explicit time integration scheme reduces the time required to compute the brain deformations by two orders of magnitude in comparison to implicit integration typically used in commercial finite element codes like ABAQUS. Our algorithms [110] described in Section 3, are also implemented on GPU for real-time computation [58] so that the entire model solution takes less than four seconds on commonly available hardware.

Unlike most purely image-based registration methods, the application of the biomechanics-based approach does not require any parameter tuning, and the results presented in the next Section demonstrate the predictive (rather than explanatory) power of this method.

### 5.1.5 Results and validation

Below we present qualitative and quantitative evaluation of our biomechanical modeling based on intraoperative MR measurements taken in Surgical Planning Laboratory at Brigham and Women's Hospital, Boston. First, we verify the plausibility of the registration results by inspecting the computed displacement vector at voxels of the preoperative image domain.

Next, to obtain a qualitative assessment of the degree of alignment after registration, we examine the overlap of corresponding anatomical features of the intraoperative and registered preoperative image. To ensure repeatability and objectivity of the analysis, we chose Canny edges [111] as features for comparison [112]. Edges are easily recognizable features and a large proportion of them correspond to boarders of real anatomical structures visible in the image. Edges can be detected using techniques that are automated and fast. Canny edges obtained from the intraoperative and registered preoperative image slices are labelled in different colors and overlaid.

For a quantitative evaluation of the accuracy of the displacement calculations we used the Edge-based Hausdorff distance. This methodology, based on pioneering work of Huttenlocher et al. [113], is described in detail in Garlapati et al. [112].

*Results*

We compared the deformation fields estimated by the biomechanical model to these obtained from the BSpline transform (available in 3D Slicer [114]) used to register pre- to intraoperative neuroimages. These deformation fields are three dimensional. However, for clarity, we display only arrows representing 2D vectors (x and y components of displacement), overlaid on preoperative (undeformed) slices, Figure 7. Each of these arrows represents the displacement of a voxel of the preoperative image domain. In general, the displacement fields calculated by the BSpline registration algorithm are similar to the predicted displacements by the biomechanical model at the outer surface of the brain, but in the interior of the brain volume the displacement vectors differ in both magnitude and direction.

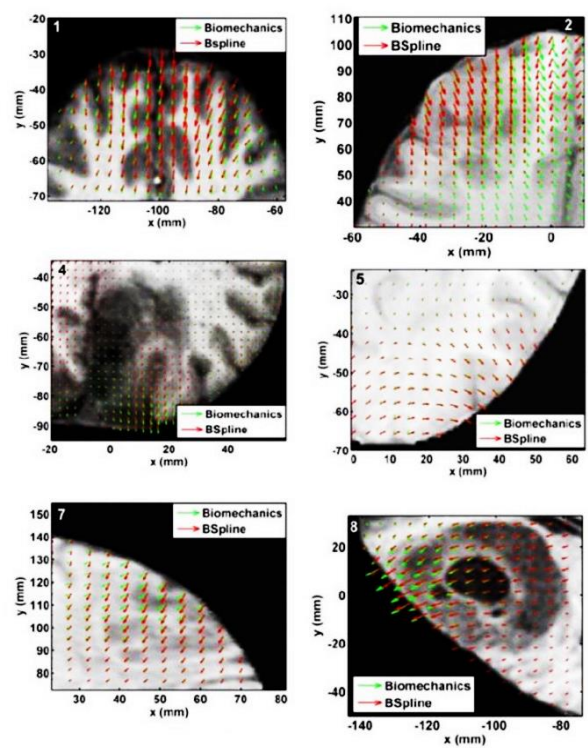

**Figure 7.** The predicted deformation fields overlaid on an axial slice of preoperative image (six examples). An arrow represents a 2D vector consisting of the x (R-L) and y (A-P) components of displacement at a voxel center. Green arrows show the deformation field predicted by the biomechanical model. Red arrows show the deformation field calculated by the BSpline algorithm. The number on each image denotes a particular neurosurgery case. Modified from Mostayed et al. [69].

From Figure 8 we can see that misalignment between the edges detected from the intraoperative images and the edges from the preoperative images updated to the intraoperative brain geometry are much lower for the biomechanics-based warping than for BSpline registration. This is an indication that the biomechanics-based prediction of the brain deformations may perform more reliably than the BSpline registration algorithm if large deformations are involved.

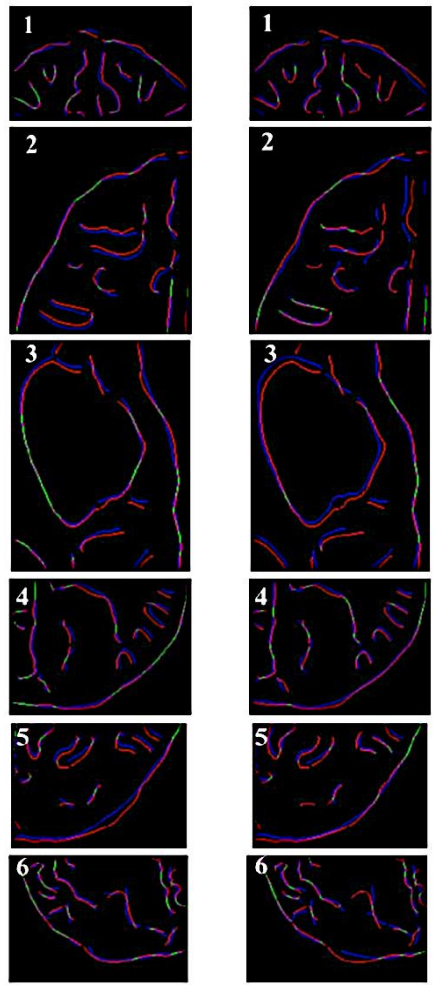

**Figure 8.** Canny edges extracted from intraoperative and the registered preoperative image slices overlaid on each other. Red lines represent the non-overlapping pixels of the intraoperative slice and blue lines represent the non-overlapping pixels of the pre-operative slice. Green lines represent the overlapping pixels. Pink color is a result of red and blue pixels overlapping. The number on each image denotes a particular neurosurgery case. For each case, the left image shows edges for the biomechanics-based warping and the right image shows edges for the BSpline-based registration. Image modified from Mostayed et al [69].

The plot of percentile edge-based Hausdorff distance (HD) versus the corresponding percentile estimates the percentage of edges whose displacements have been computed with sufficient accuracy. As the accuracy of edge detection is limited by the image resolution, an alignment error smaller than two times the original in-plane resolution of the intraoperative image (which is 0.86 mm for the cases considered) is difficult to avoid [115]. Hence, for the clinical cases analyzed here, we considered any edge pair having an HD value less than 1.7 mm to be successfully registered. This choice is consistent with the fact that it is generally accepted that manual neurosurgery has an accuracy of at best 1 mm [115, 116]. Figure 9 demonstrates that biomechanical warping was able to successfully register more edges than the BSpline registration.

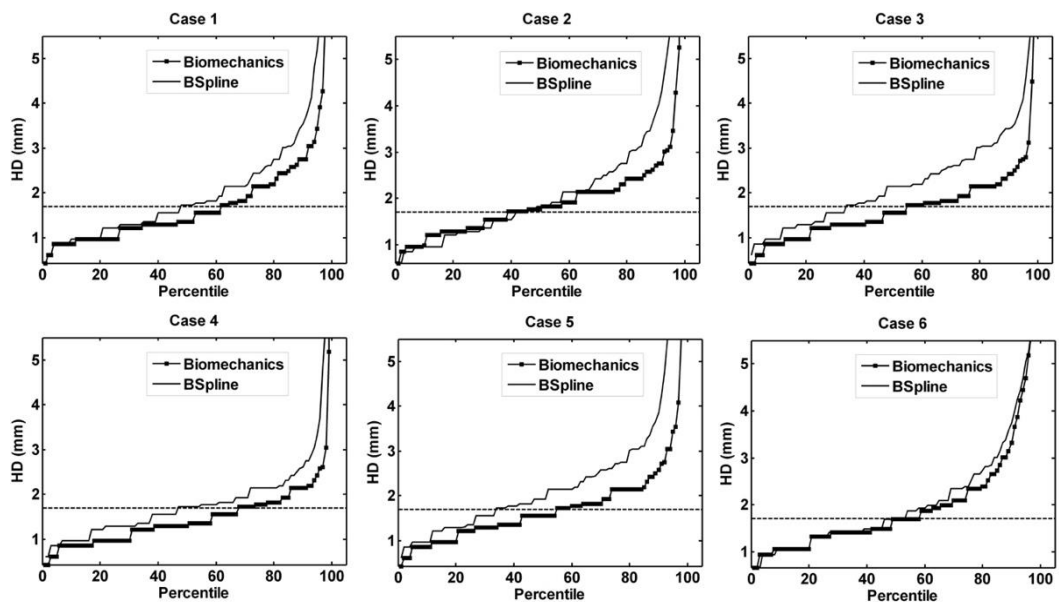

**Figure 9.** The plot of percentile edge-based Hausdorff distance between intraoperative and registered preoperative images against the corresponding percentile of edges for axial slices. The horizontal line is the 1.7 mm mark chosen as a threshold for successful registration. Six representative examples. Image modified from Mostayed et al [69].

*5.2  Craniotomy-induced brain shift simulation using a meshless method*

As both meshless and finite element methods are convergent and accurate tools to solve nonlinear partial differential equations of continuum mechanics, one expects that the brain deformations during surgery can be computed with equivalent accuracy using a meshless rather the finite element method. This is indeed the case and savings in time and effort during patient specific computational model preparation are immense as meshing is eliminated [14] and "hard" segmentation is replaced with fuzzy tissue classification [117].

In Figure 10 we present predicted, using our meshless MTLED algorithm [80], position of tumor and ventricle contours overlaid on intraoperative MRI. Solution obtained with the finite element method, for practical purposes the same as the one obtained with our meshless method, is included for comparison.

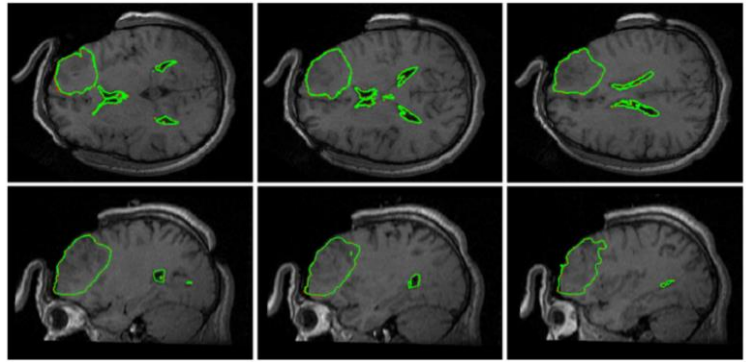

**Figure 10.** Intraoperative MRIs overlaid with contours (green lines) of the deformed tumor and ventricle surfaces as generated by our MTLED-based suite of algorithms. The three transverse sections (top row) and three sagittal sections (bottom row) were taken at 5 mm intervals. For reference, the red lines represent the contours of the deformed ventricles and tumor computed by FEM, but these are almost entirely obscured by the very similar MTLED results.

In Figure 11 we show ventricular surface color-coded with the differences between computed and observed positions of points on this surface. The predominantly blue color demonstrates that our computed displacements are accurate to approximately 0.5 mm.

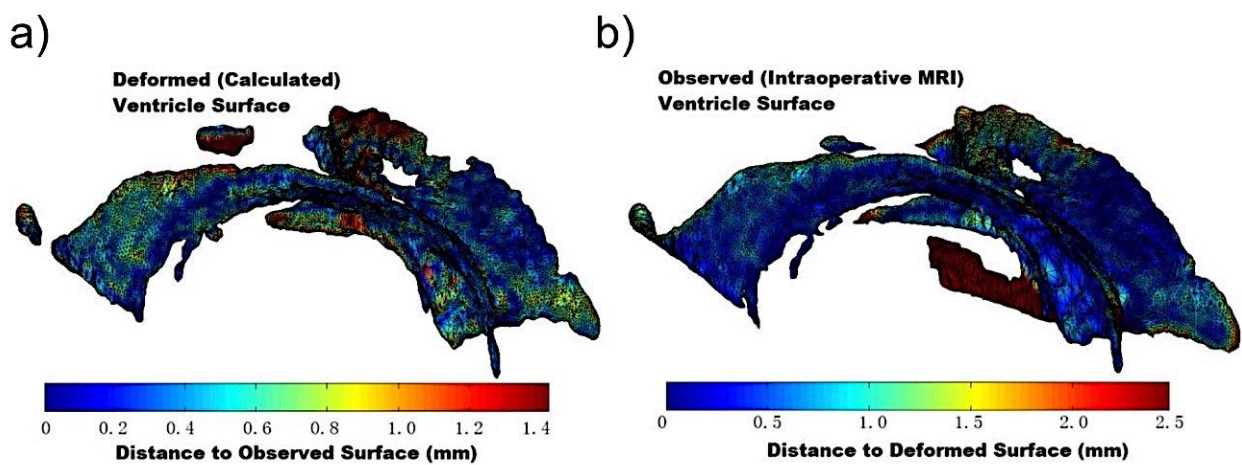

**Figure 11**. Differences between calculated (using our MTLED algorithm, Surface A) and observed with intraoperative MRI (Surface B) ventricle surfaces. a) Surface A. The colors represent the distance d(a,B) from point a on Surface A to the nearest point on Surface B generated from the segmented intraoperative MRIs. The scale reaches 95% Hausdorff Distance H95(A,B); b) Surface B. Colors representing the distance d(b,A). Image modified from Miller et al [118].

Some researchers in the field of numerical solid mechanics believe that meshless methods, while convenient for model generation, are uncompetitive in computational efficiency due to a high cost (as compared to the finite element method) of shape function computation. The simulation presented here proves that this is not the case. Our calculations were performed on a machine with an Intel Core i7 930 2.8 GHz processor and 4 GB of physical memory. The calculation time of our fully nonlinear problem with almost 100,000 degrees of freedom was 19.2 s for the approximately 1000 time steps required to obtain convergence. This is certainly fast enough for the method to be useful in intraoperative applications. Such an excellent computational efficiency is due to the fully explicit nature of our MTLED algorithm and the use of Total Lagrangian Formulation allowing precomputation of spatial derivatives.

*5.3 Electrode localization in epilepsy surgery*

Biomechanics-based prediction of brain deformations promises significant benefit in surgical treatment of epilepsy, a chronic neurological disorder. Surgical intervention can be curative, but is "arguably the most underutilized of all proven effective therapeutic interventions in the field of medicine" [119].

Intracranial EEG (iEEG) is the clinical gold standard for functional localization of the seizure onset zones (SOZ); invasive electrodes are implanted and monitored for several days, and then removed during a second surgery when the resection is performed [120]. Collecting data from directly on or within the cortex significantly increases the spatial data fidelity as compared to scalp EEG measurements, and allows more accurate identification of the SOZ. Final placement of electrodes must be verified to enable clinicians to associate measured data with preoperative images of cortical structures.

Despite the advantages of iEEG, current methods for aligning invasive electrodes with cortical structure prevent rapid, consistent clinical interpretation of measured data. Current practice of clinical electrode alignment uses a combination of intraoperative photographs and diagrams to estimate the cortical placements of electrode grids and depth electrodes (Figure 12). While this allows generalized alignment, it is not quantitative and does not provide the degree of accuracy necessary for emerging precision surgical techniques such as radio frequency (RF) or laser ablation [121, 122], and focused ultrasound [123]. To fully utilize the precision of these emerging techniques, accurate alignment algorithms proposed in this application are needed to enable precise identification of the SOZ with respect to both pre-operative and intraoperative imaging.

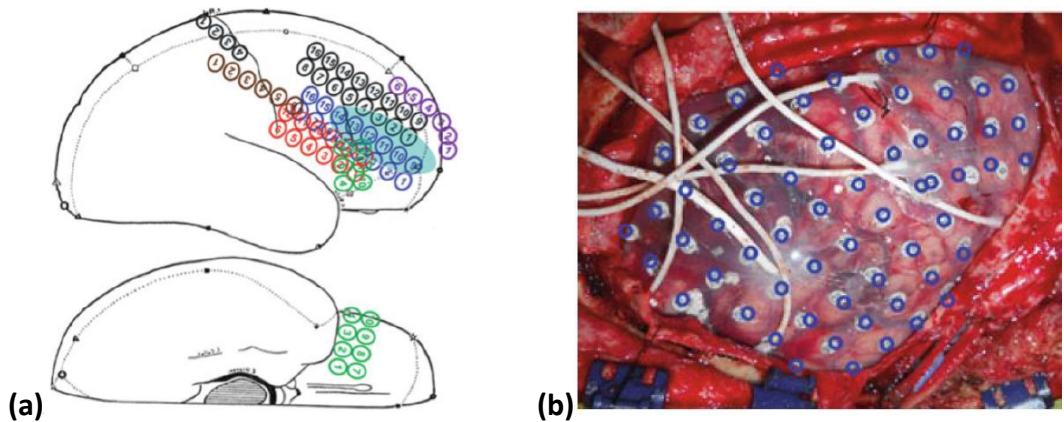

**Figure 12.** a) Surgical Diagram and b) intraoperative photographs are the current clinical standard for identifying electrode placements. Additionally, electrode positions identified from medical imaging are projected on the brain surface and shown as blue rings [124].

A significant factor impacting electrode alignment error is the physical shifting of brain tissue during invasive measurements, Figure 13. Electrode grids in the intracranial space, and the body's inflammatory response to the craniotomy, displace and deform the brain from the configuration observed in presurgical MR imaging [124-126]. To ensure accurate alignment of electrode placements and correct clinical evaluation of invasive data, brain shift must be accurately modeled and accounted for. We can address this critical problem by modeling the deformation of brain tissue during invasive electrode monitoring studies.

The biomechanical model is constructed based on the high resolution preoperative MRI. The load is defined as imposed displacements on the model surface. In intraoperative CT (Figure13c) the implanted electrodes and the grid, which define the deformed surface of the brain, are clearly visible. Widely used rigid alignment of preoperative MRI with intraoperative CT and projection of the electrode positions from the deformed brain surface (seen on CT) onto an undeformed brain surface (seen on MRI) allows the determination of surface displacements [124]. The computed deformation field is used to warp the preoperative MRI image, as shown in Figure 13d, which permits a precise localization of the seizure onset zones (SOZ) within the brain.

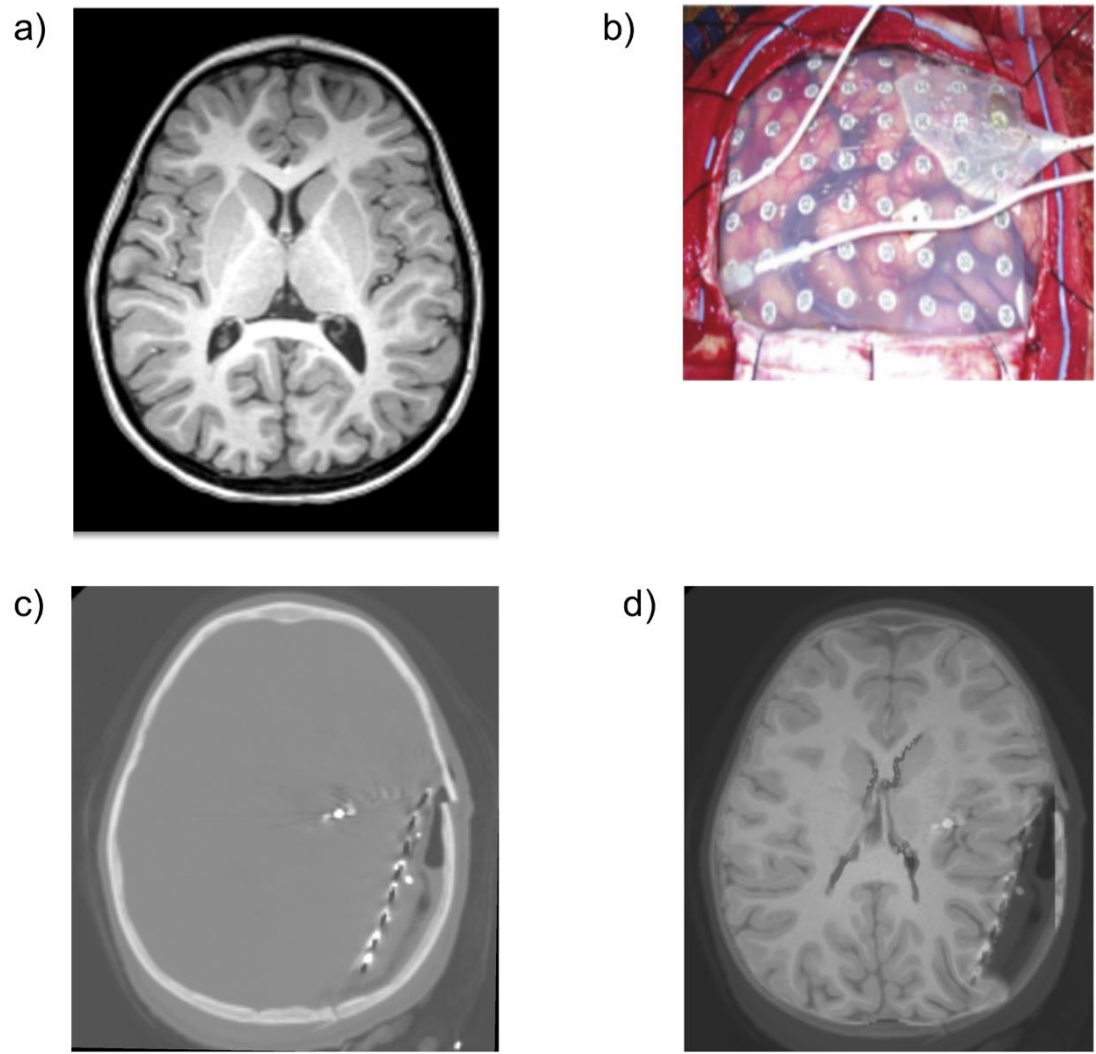

**Figure 13.** Preoperative MRI (a), intracranial electrodes inserted (b), intra-operative CT with electrodes implanted (c), and preoperative MRI registered onto intra-operative CT (d). These are 2D sections of 3D image volumes. We have successfully registered four cases [127].

## 6   Discussion and conclusions

In this review article, we have discussed modeling approaches to two applications of clinical relevance: surgical simulation and neuroimage registration. These problems can be reasonably characterized with the use of purely mechanical terms such as displacements, strains, internal forces, stresses, etc. Therefore, they can be analyzed using the methods of continuum mechanics. Moreover similar methods may find applications in modeling the development of structural diseases of the brain [128-133].

As the brain undergoes large displacements (~10 – 20 mm in the case of a craniotomy-induced brain shift and even more as a result of invasive electrode placement) and its mechanical response to external loading is strongly nonlinear, we recommend the use of general, nonlinear procedures for the numerical solution of the proposed models. We strongly believe that the use of linear models, based on the assumption of infinitesimal deformations, should be avoided.

The brain's complicated mechanical behavior: nonlinear stress-strain, stress-strain-rate relationships and much lower stiffness in extension than in compression require very careful selection of the constitutive model for a given application. The selection of the constitutive model for surgical simulation problems depends on the characteristic strain-rate of the process to be modeled and to a certain extent on computational efficiency considerations. Fortunately, as shown in [26, 106] the precise knowledge of patient-specific mechanical properties of brain tissue is not required for intraoperative image registration, which can be formulated as a Dirichlet-type problem.

Satisfying the requirement of real time computations for the model sizes (minimum of around 40,000 degrees of freedom) recommended in Table 1 is in practice impossible without parallel computing. The solution methods (Total Lagrangian Explicit Dynamics TLED finite element and Meshless Total Explicit Lagrangian algorithms), we applied to obtain the results reported in this article, rely on explicit time integration. Such methods have very modest internal memory requirements as they do not involve solving systems of algebraic equations and do not require iterations even for highly nonlinear problems. This makes the solutions methods that employ explicit time stepping inherently data-parallel, and therefore particularly suitable for multithreading. We implemented in parallel both TLED and MTLED algorithms. The implementation on GPU (see Section 5) satisfies the real time computation requirements with the entire model solution of under four seconds (around 1 ms for each computation step). It should be noted, however, that even for the algorithms that do not involve solving systems of equations and despite rapid progress in the specialized programming environments dedicated to GPU (such as NVIDIA's Compute Unified Device Architecture (CUDA) [134] and PyCUDA [135], and Open Computing Language OpenCL [136]), implementation of computational biomechanics algorithms on GPU remains a non-trivial programming task that may require substantial time and effort. For applications in image registration, utilization of hyperthreading features available in CPUs installed in many off-the-shelf personal desktop and laptop computers may provide sufficient computation speed.

Computational mechanics has become an empowering discipline that has led to greater understanding and advances in modern science, technology and engineering [137]. One of the greatest challenges for the field of Computational Mechanics is to translate it's phenomenal success to medicine. Computational Mechanics can become a central enabling discipline contributing to the creation of a new era of personalized medicine based on patient-specific scientific computations. Computer-Integrated Surgical systems (CIS), based on patient-specific scientific computations, will enable a surgeon to use sophisticated (but robust!) computational tools for diagnosis, prognosis and treatment planning as well as to simulate surgery within the operating theatre in real time using readily-available computing facilities, and to visualize the results immediately.

However, before this vision can be realized and Computer-Integrated Surgery systems based on computational biomechanical models can become as widely used as Computer-Integrated Manufacturing systems are now, a number of challenges must be met. As we deal with individual patients, methods to produce patient-specific computational grids quickly and reliably must be improved [61]. Substantial progress in automatic meshing methods is required, or alternatively meshless methods may provide a solution. Computational efficiency is an important issue, as intraoperative applications, requiring reliable results within approximately 40 seconds, are most appealing. Progress can be made in nonlinear algorithms by identifying parts that can be precomputed, and parts that do not have to be calculated at every time step. One such possibility is to use the Total Lagrangian Formulation of continuum mechanics [54, 138, 139], where all field variables are related to the original (known) configuration of the system and therefore most spatial derivatives can be calculated during the preprocessing, before the time-critical simulation commences. Implementation of algorithms in parallel on networks of processors, and harnessing the computational power of graphics processing units [58, 140, 141] provide a challenge for coming years.

**Acknowledgements**

The funding from the Australian Government through the Australian Research Council ARC (Discovery Project Grants DP160100714, DP1092893, and DP120100402) and National Health and Medical Research Council NHMRC (Project Grants APP1006031 and APP1144519) is greatly acknowledged. We thank the Raine Medical Research Foundation for funding G. R. Joldes through a Raine Priming Grant, and the Department of Health, Western Australia, for funding G. R. Joldes through a Merit Award. This investigation was also supported in part by NIH grants R01 NS079788, R01 EB019483, R42 MH086984, P41 EB015902, P41 EB015898, U24 CA180918 and by a research grant from the Boston Children's Hospital Translational Research Program.